\documentclass[12pt,a4paper,showkeys,prl]{revtex4}

\usepackage{graphicx}
\usepackage{amsmath}
\usepackage{epstopdf}
\usepackage{amsbsy}
\usepackage{bm}
\newcommand{\down}{\downarrow}
\newcommand{\up}{\uparrow}

\begin{document}

\title{Modulations of the local pairing interaction near magnetic impurities in d-wave superconductors}
\author{A. T. R\o mer$^{1*}$, S. Graser$^{2}$, P. J. Hirschfeld$^{3}$, and B. M. Andersen$^1$}
\affiliation{$^1$Niels Bohr Institute, University of Copenhagen, Universitetsparken 5, DK-2100 Copenhagen,
Denmark, $^*$ ar@fys.ku.dk, Tlf.: +45 35320451, Fax: +45 35320460\\
$^2$Theoretical Physics III, Center for Electronic Correlations and Magnetism, Institute
of Physics, University of Augsburg, D-86135 Augsburg, Germany\\
$^3$Department of Physics, University of Florida, Gainesville, Florida 32611, USA}

\begin{abstract}
The spin-fluctuation based pairing mechanism has proven successful in explaining the pairing symmetries due to Fermi surface nesting of both cuprates and iron-based materials. In this work, we study signatures of a spin-fluctuation mediated pairing at the local scale. Specifically, we focus on magnetic impurities and calculate both the local antiferromagnetism and the resulting modulated pairing interaction. The latter gives rise to distinct local enhancements of the superconducting gap in the immediate vicinity of the impurities. Our results show that Coulomb-driven pairing naturally leads to unusual superconducting gap modulations near disorder potentials.    
\end{abstract}


\date{\today}
\maketitle

\section{Introduction}

In the cuprate Bi$_2$Sr$_2$CaCu$_2$O$_{8+\delta}$ a correlation between oxygen dopant positions and local enhancements of the superconducting gap has been observed by STM measurements~\cite{McElroy05,Hofmann12}. This correlation was explained at a phenomenological level by introducing a spatially modulated pairing strength which is enhanced at the impurity sites~\cite{Nunner05}. More generally, the idea of a spatially modulated pairing interaction caused by the combination of disorder and electronic correlations have been somewhat successful in explaining a series of experiments \cite{Nunner05,Nunner06,Nunner07,Slezak07,Andersen07}. The microscopic origin of the modulated pairing potential remains, however, unsettled despite several
recent discussions \cite{maska07,foyevtsova09,johnston10,khaliullin10,christensen11,wei11}. We have recently shown that such local enhancements of the pairing potential and hence the observed superconducting gap is a natural consequence of spin-fluctuation mediated pairing \cite{Romer12}. For an overview of spin-fluctuation mediated pairing in cuprates and iron-pnictides we refer to Refs. \onlinecite{Scalapino95,Hirschfeld11}.
Here, we focus on the role of magnetic impurities, and study their effects on the pairing potential in $d$-wave superconductors. We utilize the recently developed real-space formulation of the effective pairing interaction in real-space. Similar to the situation of non-magnetic disorder \cite{Romer12}, it is found that magnetic impurities may lead to significant local enhancements of the superconducting gap.

\section{Model}
In the initial step of the calculation of the effective real-space pairing potential, we obtain the spin-resolved charge densities calculated self-consistently in the normal state using a mean-field approximation to the one-band Hubbard model
\begin{equation}
H_0=\sum_{i,j, \sigma}t_{i,j}c_{i\sigma}^{\dagger}c_{j\sigma}+ \sum_{i \sigma} (U \langle n_{i\sigma} \rangle - \mu ) n_{i\bar\sigma} 
+ \sum_{i \sigma} V_{\rm imp} \delta(r_i-r_{i_{\rm imp}}) n_{i\sigma}.
\label{eq:H}
\end{equation}
Here, $c_{i\sigma}^{\dagger}$ refers to creation of an electron with spin $\sigma$ at lattice site $i$, and $n_{i\sigma}$ is the number operator of spin $\sigma$ particles at site $i$. 
Note that this Hamiltonian also contains the impurity potential $ V_{\rm imp}$ at a site placed at position $r_{i_{\rm imp}}$. A brute force diagonalization of Eq.(\ref{eq:H}) allows us to obtain the effective interaction $V_{\text{eff}}(i,j)$ which is due to longitudinal and transverse spin fluctuations. Using the approach of Berk and Schrieffer \cite{berk:1966}, we obtain a real-space formulation of the effective pairing interaction which can be written as \cite{Romer12}
\begin{equation}
 V_{\text{eff}}(i,j)=U+\frac{U^3\chi_{\down\down}\chi_{\up\up}}{\hat{1}-U^2\chi_{\down\down}\chi_{\up\up}}\Big|_{(i,j)}+\frac{U^2\chi_{\down\up}}{\hat{1}-U\chi_{\down\up}}\Big|_{(i,j)}.
\label{eq:Veff}
\end{equation}
The susceptibilities entering Eq.(\ref{eq:Veff}) are real-space matrices given by
\begin{equation}
\chi_{ij}^{\sigma \sigma'}\!(\omega\!=\!0)\!=\!\sum_{m,k}u_{mi\sigma}u_{mj\sigma}u_{kj\sigma'}u_{ki\sigma'} \frac{f(E_{m,\sigma})-f(E_{k,\sigma'})}{E_{k\sigma'}-E_{m\sigma}+i\eta},
\end{equation}
in terms of the eigenvectors $u_{m\sigma}$ and eigenvalues $E_{m\sigma}$ obtained in the diagonalization of Eq.(\ref{eq:H}).

After the calculation of the effective spin-fluctuation mediated pairing, the density and superconducting gap values are calculated self-consistently using a mean-field approach for both density and pairing channels
\begin{equation}
H_{\rm SC}=
H_0+\sum_{i,j}\left[\frac{\Delta_{i,j}}{2}(c_{i\up}^\dagger c_{j\down}^\dagger-c_{i\down}^\dagger c_{j\up}^\dagger)+h.c.\right],
\label{eq:Hdelta}
\end{equation}
where $\Delta_{i,j}=\frac{V_{\text{eff}}(i,j)}{2}\langle c_{j\down} c_{i\up}-c_{j\up} c_{i\down} \rangle$. The factor $1/2$ arises from the restriction to the singlet pairing channel. \\
In this real-space study, we investigate the impurity effects on the superconducting gap which arise due to a spin-fluctuation mediated pairing mechanism \cite{Romer12}. The self-consistent approach allows a detailed study of the co-existence of spin density variations and superconducting gap modulations. We will focus on the effect of point-like magnetic impurities.
 
\begin{figure}[h!]
 \begin{center}
\includegraphics[width=15cm]{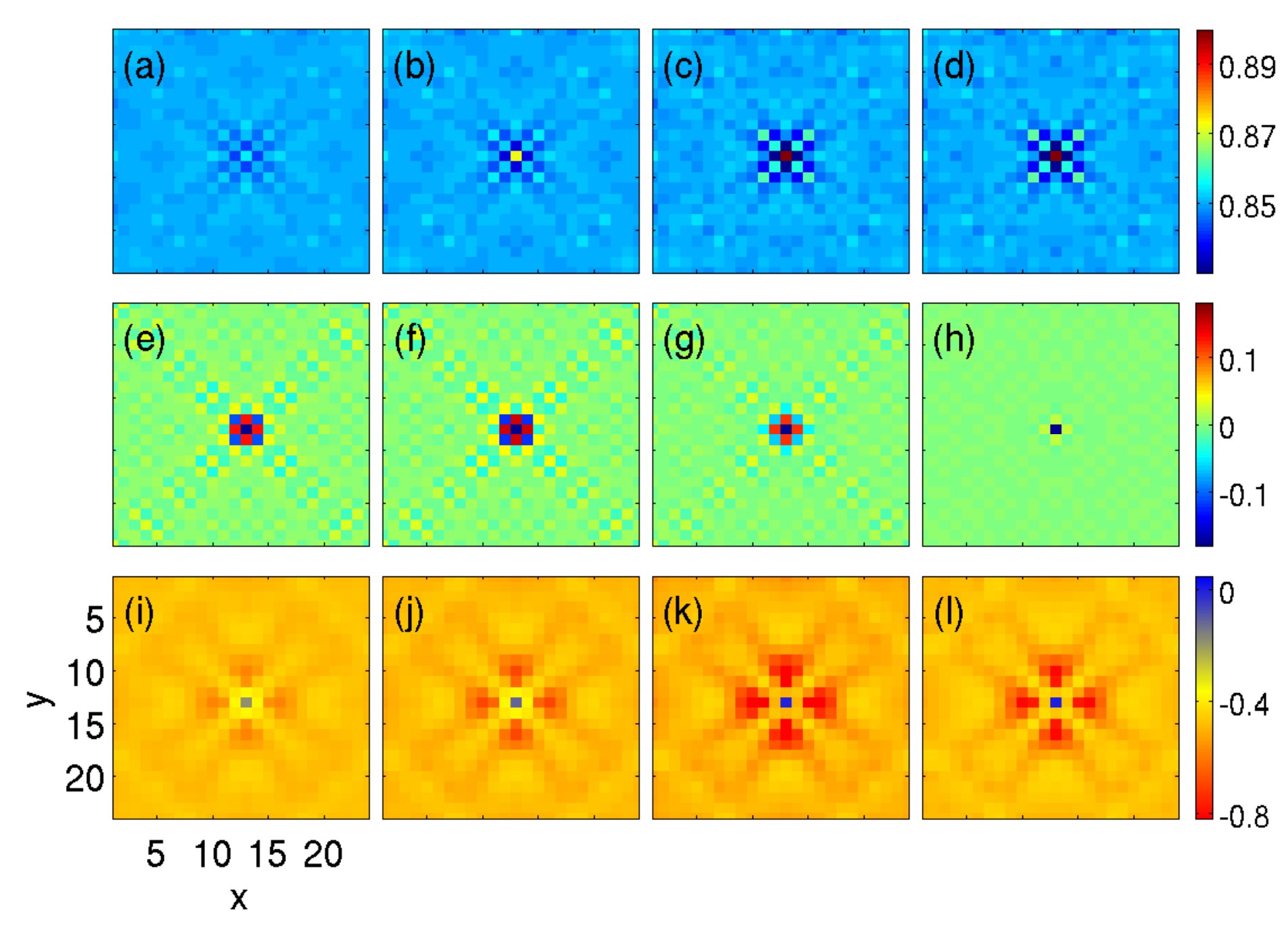}
 \end{center}
 \caption{(a-d) Local charge density of the normal phase. An impurity is positioned at site $r_{i_{\rm imp}}=(13,13)$ and the impurity strengths are in units of the nearest neighbor hopping constant $t$: (a) $V_{\rm imp}=0.5$, (b) $V_{\rm imp}=1$, (c) $V_{\rm imp}=10$ and (d) $V_{\rm imp}=100$. For the results presented here the system size is $24 \times 24$ and the parameters are: $U=2.2$, $t'=-0.3$ in units of $t$, and the doping is $x=0.15$. In panels (e-h) we show the magnetization, and in (i-l) the local pairing interaction between nearest neighbors, for the same parameters as in (a-d).}
\label{fig1}
\end{figure}

\section{Results and Discussion}
Prior to the calculation of the effective pairing interaction, the spin densities are calculated self-consistently for a system containing a single point-like magnetic impurity.   
In Fig.~\ref{fig1} (a-d) the total electron density  $\rho_i=\langle n_{i\up}\rangle+\langle n_{i\down}\rangle$ is shown as a function of lattice site. Local density variations occur close to the impurity site, and a small increased charge density is evident at the impurity site caused by a finite magnetization. In Fig.~\ref{fig1} (e-h) we show the induced magnetization, $m_i=\langle n_{i\up}\rangle-\langle n_{i\down}\rangle$, as a function of lattice site. Local antiferromagnetism is induced around the impurity, and is most extended because of smaller magnetization induced by weaker impurity potentials.\\ 
The effective superconducting pairing interaction is shown in Fig.~\ref{fig1} (i-l). The interaction is calculated in real-space from the expression given in Eq.~(\ref{eq:Veff}). Interestingly, the effect of the impurity is not only to suppress the interaction at the impurity site. In fact, we see an enhancement of the pairing interaction at sites around the impurity. The enhancement effect is related to a re-distribution of spin densities which takes place in the presence of an impurity as reflected in the total density modulations and local magnetization \cite{Romer12}. The effect is most pronounced in the two strongest magnetic impurity cases, see Fig.~\ref{fig1} (k,l). \\
In the limit of very strong magnetic impurities, see Fig.~\ref{fig1}(h), local antiferromagnetism is essentially confined to the impurity site, and still we find significant local gap enhancement as seen from Fig.~\ref{fig1}(l). It is also known that non-magnetic impurities cause a local enhancement of the pairing interaction even though no local antiferromagnetism is induced in the normal phase of these systems \cite{Romer12}. It therefore turns out that 
the enhancement effect is not dependent on induced antiferromagnetism and it is local variations of spin densities rather than a difference between spin densities that causes the effect.\\
\begin{figure}[h!]
 \begin{center}
\includegraphics[width=15cm]{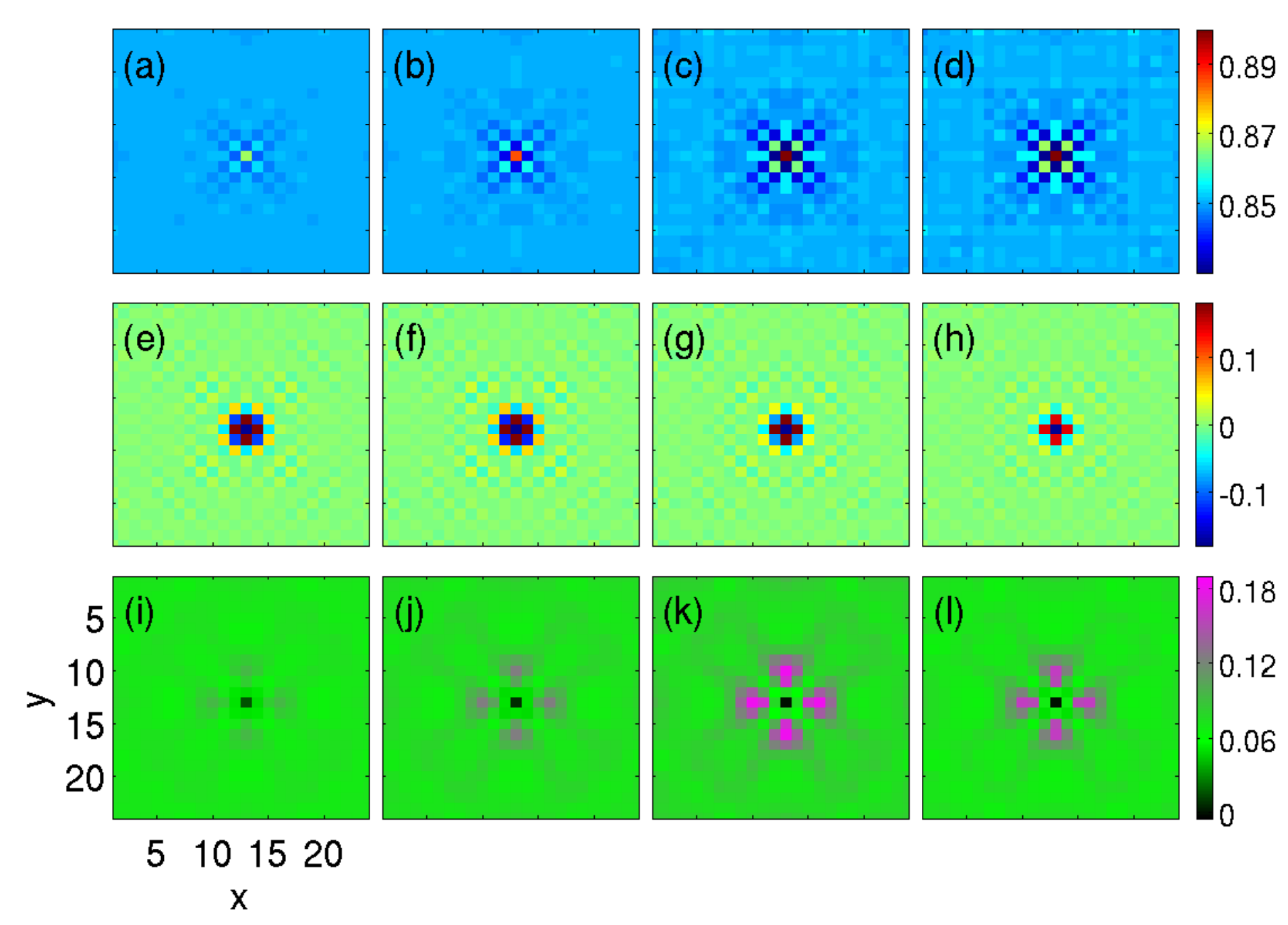}
 \end{center}
 \caption{(a-d) Local charge density in the superconducting phase. An impurity is positioned at site $r_{i_{\rm imp}}=(13,13)$ and the impurity strengths are in units of the nearest neighbor hopping constant $t$: (a) $V_{\rm imp}=0.5$, (b) $V_{\rm imp}=1$, (c) $V_{\rm imp}=10$ and (d): $V_{\rm imp}=100$. For the results presented here the system size is $24 \times 24$ and the parameters are: $U=2.2$, $t'=-0.3$ in units of $t$ and doping is $x=0.15$. (e-h) Local magnetization for the same impurities and parameters as (a-d). (i-l) Magnitude of the local $d$-wave superconducting order parameter $\Delta(i)=\frac{1}{2}[\Delta_x(i)-\Delta_y(i)]$ for the same impurities and parameters as (a-d).}
\label{fig2}
\end{figure}
Including the effective pairing interaction shown in Fig.~\ref{fig1} (i-l), we calculate the superconducting gap self-consistently by diagonalization of the mean-field Hamiltonian given in Eq.~(\ref{eq:Hdelta}). As apparent from Fig.~\ref{fig2} (a-h), the total density arrangement around the impurity is roughly unchanged while more antiferromagnetism is induced around the strongest impurity. The superconducting $d$-wave order parameter is locally enhanced, see Fig.~\ref{fig2} (i-l), and the real-space structure resembles that of the local pairing potential. The enhancement of the superconducting gap thus occurs at the borders of regions with spin density modulations and the enhancement effect competes locally with antiferromagnetic order. It appears that there exists some optimal intermediate impurity strength at which the gap enhancement becomes strongest, Fig.~\ref{fig2} (k).

\begin{figure}[h!]
 \begin{center}
\includegraphics[width=7.5cm]{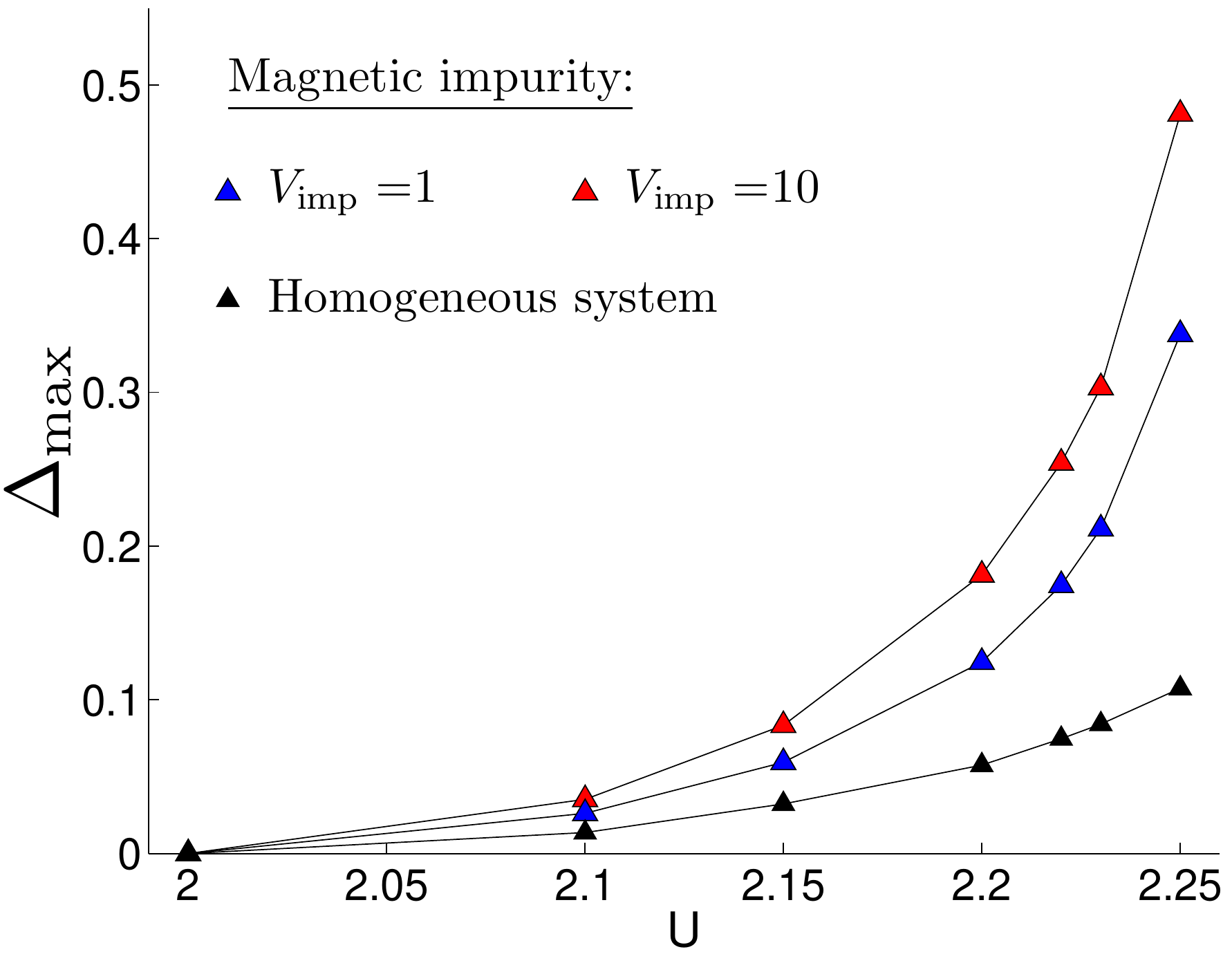}
\includegraphics[width=7.5cm]{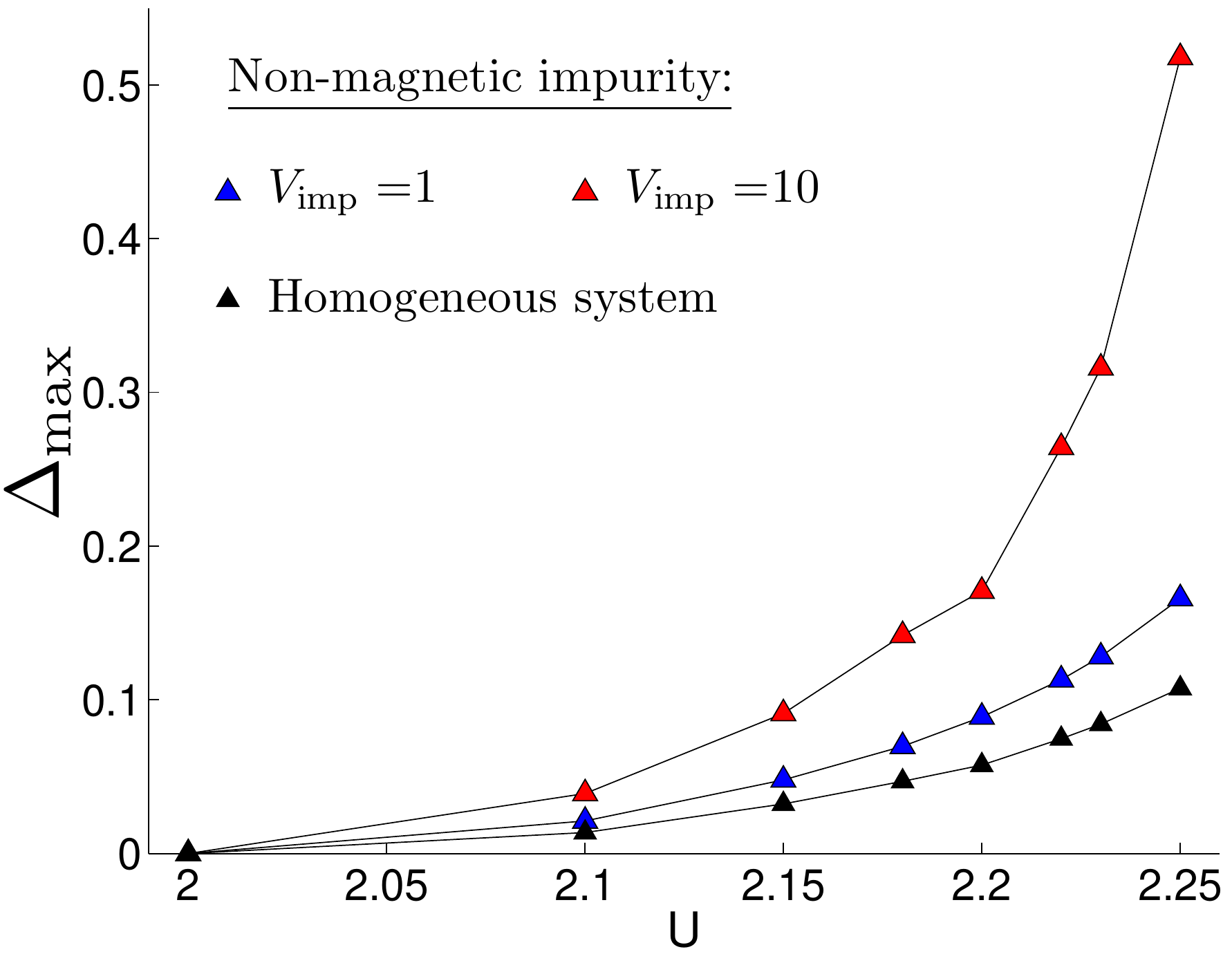}
 \end{center}
 \caption{Left: Maximum superconducting gap value as a function of Coulomb interaction strength, $U$, for point-like magnetic impurities of strengths $V_{\rm imp}=t, 10t$. Also the gap value in the homogeneous system is shown for reference. Right: Maximum superconducting gap value as a function of Coulomb interaction strength, $U$, for point-like non-magnetic impurities of strengths $V_{\rm imp}=t, 10t$.}
\label{fig3}
\end{figure}
In Fig.~\ref{fig3} the maximum value of the superconducting gap is shown as a function of Coulomb interaction strength, $U$, for a magnetic point-like impurity (left) and a non-magnetic point-like impurity (right) \cite{Romer12}. In the case of a non-magnetic impurity, no local antiferromagnetism is induced in the vicinity of the impurity. However, the enhancement effect is still present and the real-space structure of the enhancement resembles that of the magnetic impurity, shown in Fig.~\ref{fig2} (i-l). One interesting difference occurs for the weak impurity, where a magnetic impurity leads to a stronger enhancement than the non-magnetic impurity, compare the blue data points in Fig.~\ref{fig3}.\\ 
In Fig.~\ref{fig3} the effect of tuning $U$ becomes apparent. Even though small changes in $U$ will not significantly alter the spin density at each lattice site it has a great impact in the RPA-like pairing interaction. By increasing the Coulomb interaction we approach the singularities in Eq.~(\ref{eq:Veff}) and therefore the "Stoner enhancement" of the superconducting gap is tuned by $U$. Thus, the Coulomb interaction governs both the local gap enhancements and the development of antiferromagnetism, and the final superconducting gap results from a balance between them.

\section{Conclusions}
We have shown that a local enhancement of the superconducting gap occurs in the vicinity of a magnetic impurity if the pairing interaction is mediated by spin-fluctuations. The real-space structure of the gap enhancement is robust to the impurity strength and resembles that of a point-like non-magnetic impurity. The effect arises due to local variations in the spin densities caused by the impurity. Due to these variations, the spin susceptibilities become inhomogeneous thereby enabling local enhancements of the effective pairing interaction. The effect is apparent already from the bare susceptibility\cite{Romer12}, but is enhanced by correlations within RPA. 

\section{Acknowledgement}

B.M.A. acknowledges support from The Danish Council for Independent Research $|$ Natural Sciences. P.J.H acknowledges support from NSF-DMR-1005625.

\end{document}